\documentclass[aps,showpacs,manuscript,12pt]{revtex4}
\usepackage{amssymb}
\usepackage{amsmath}
\usepackage{graphicx}

\begin{document}
\title{\textbf{IKT-approach to MHD turbulence$^{\S }$}}
\author{M. Tessarotto$^{a,b}$} \affiliation{\ $^{a}$Department of Mathematics
and Informatics, University of Trieste, Italy\\ $^{b}$Consortium
of Magneto-fluid-dynamics, University of Trieste, Italy}
\begin{abstract}

An open issue in turbulence theory is related to the determination
of the exact evolution equation for the probability density
associated to the relevant (stochastic) fluid fields. Such an
equation in the usual approaches to turbulence reproduces, at most
in an approximate sense, the correct fluid equations. In this
paper we present a statistical model which applies to an
incompressible, resistive and quasi-neutral magnetofluid. The
approach is based on the formulation of an inverse kinetic theory
(IKT) for the full set of MHD equations appropriate for an
incompressible, viscous, quasi-neutral, isentropic, isothermal and
resistive magnetofluid. Basic feature of the new approach is that
it relies on first principles - including in particular the exact
validity of the fluid equations - and thus permits the
determination of the correct evolution equation for the
probability density. Specific application of the theory here
considered concerns the case of statistically homogeneous and
stationary MHD turbulence.
\end{abstract}
\maketitle

%$^{\star }$} \url{http://cmfd.univ.trieste.it}.

%%%%%%%%%%%%%%%%%%%%%%%%%%%%%%%%%%%%%%%%%%%%
%% MAINMATTER
%%%%%%%%%%%%%%%%%%%%%%%%%%%%%%%%%%%%%%%%%%%%

\section{Introduction}

The goal of this paper is to investigate a well-known theoretical
issue of fluid dynamics. This is concerned with the rigorous
formulation of the so-called \emph{turbulence problem}, i.e., the
description of the dynamics of a classical fluid in the presence
of stochastic sources. We refer here, in particular, to the
statistical description of turbulence for incompressible fluids
based on its phase-space approach, namely achieved by representing
the turbulent fluid means of a suitable phase-space probability
density function (pdf). In such a context, an open issue is
related to the determination of the exact evolution equation for
the probability density associated to the relevant (stochastic)
fluid fields in an incompressible turbulent fluid. It is well
known, in fact, that in customary approaches to turbulence (see
for example Monin and Yaglom \cite{Monin1975} 1975 and Pope, 2000
\cite{Pope2000}) this equation reproduces only in an approximate
sense the correct fluid equations. In this paper we present a
statistical theory which applies to an incompressible, resistive
and quasi-neutral magnetofluid.

The purpose of this paper is to propose a solution based on the
formulation of an inverse kinetic theory (IKT) for an
incompressible, viscous, quasi-neutral, isentropic and resistive
magnetofluid. Basic feature of the theory is that it permits the
determination of the exact evolution equation for the probability
density.

% ---------------------------- The origin -------------------------------

\subsection{Deterministic and stochastic descriptions of fluids}

In fluid dynamics the state of the fluid is assumed to be
prescribed by an appropriate set of suitably smooth functions
$\left\{ Z\right\} \equiv \left\{ Z_{i},i=1,n\right\} $ denoted as
\emph{fluid fields}$.$ These are required to be real functions
which are at least continuous in all points of
a closed set $\overline{\Omega }\times \overline{I},$ closure of the {%
extended configuration domain }$\Omega \times ${$I.$ \ In the
remainder we shall require that:}

\begin{enumerate}
\item {$\Omega${\ (configuration domain) is a bounded subset of the
Euclidean space }$E^{3}$ on $R^{3};$}

\item $I$ is identified, when appropriate, either with a bounded time
interval, \textit{i.e.}, $I${$=$}$\left] {t_{0},t_{1}}\right[
\subseteq \mathbb{R},$ or with the same time axis ($\mathbb{R}$);

\item in the open set $\Omega \times ${$I$} the functions $\left\{ Z\right\}
,$ are assumed to be solutions of an appropriate closed set of
PDE's, denoted as \emph{fluid equations;}

\item by assumption, these equations together with appropriate initial and
boundary conditions are required to define a well-posed problem
with unique strong solution defined everywhere in $\Omega \times
${$I$}.
\end{enumerate}

Depending on the definition of initial conditions (A), boundary
conditions (B), possible volume forces (C), as well as, more
generally, also the functional form assumed for the fluid
equations (D), two types of descriptions of the fluid,
\textit{deterministic} and \textit{stochastic}, can in principle
be distinguished, in which the fluid fields are treated
respectively as deterministic or stochastic functions (see
definitions in Appendix). \ The two descriptions are obtained,
respectively, when all conditions (A), (B), (C) and (D) are
deterministic, or at least one of them [A,B, C or D] is prescribed
in a stochastic sense (see below). For definiteness, in the
following (except for Sec.7) we shall formally rule out case D,
imposing that the functional form of the fluid equations remains
the same in both cases.

\subsection{Two types of turbulence - Homogeneous and stationary turbulence}

The stochastic description of fluids is usually denoted as \textit{turbulence%
} \cite{Theory}. In this respect, however, two different types of
phenomena should be distinguished, \textit{i.e.},

\begin{itemize}
\item \textit{physical turbulence}

\item \textit{numerical turbulence}.
\end{itemize}

The first one arises due to "physical effects", namely the
specific definition adopted for the fluid equations and for the
related initial and boundary conditions (these definitions may be
considered, in a sense, as physical characteristics of a fluid,
i.e., as related to the properties of the fluid). Thus, physical
turbulence may appear only as a result of the \textit{stochastic
"sources"} indicated above. The first two (A and B) are due to
possible stochastic initial and boundary conditions. In fact, the
initial and/or boundary state of the fluid may not be known
deterministically, as implied by measurement errors, or by the
action of some type of initial and/or boundary volume/surface
stochastic forcing. Finally (case D), it is obvious that the fluid
equations for the stochastic fluids fields (\emph{stochastic fluid
equations}) may, in principle, differ in other ways (besides C),
from the corresponding deterministic fluid equations. Their
precise prescription is therefore essential for the definition of
specific turbulence models.

Instead, numerical turbulence occurs only in numerical simulations
and is due specifically to approximations involved in the
numerical solution method (for the fluid equations). Hence, it is
necessarily related to the specific algorithms adopted for the
construction of the numerical solutions.

Both types of turbulence may produce, in principle, "real" effects
to be observed in numerical experiments. However, it should be
stressed that, in practice, the distinction between the two
phenomena may be difficult, or even impossible, to ascertain in
numerical simulations. Nevertheless, from theoretical perspective
it is obvious that the distinction between the two phenomena
should be made.

This paper deals with exact properties and implications of the
fluid equations (and the related initial and boundary conditions),
disregarding completely issue related to the possible choice of
the approximate solution methods adopted for their numerical
solution. Hence, we will face only the first type of phenomena.
For this reason in the following the term "turbulence" will be
meant, in a proper sense, only in the physical sense here
specified.

Turbulence is therefore characterized by the presence of
\emph{stochastic
fluid fields }$Z_{i}\in \left\{ Z\right\} ,$ for $i=1,n$\emph{\ } (with $%
\left\{ Z\right\} $ defining the state of the fluid), which are
assumed to depend on suitable \emph{hidden variables}
$\mathbf{\alpha }=\left\{ \alpha _{i},i=1,k\right\} \in V_{\alpha
}\subseteq \mathbb{R}^{k},$ with $k\geq 1$, which are assumed
independent of $(\mathbf{r,}t)$. Hidden variables by assumption
cannot be known deterministically (in other words, in the context
of classical mechanics, they are not "observables"), and hence are
necessarily stochastic. This means that by assumption: 1) the
hidden variables $\mathbf{\alpha }$ are characterized by a
probability density $g,$ denoted as \emph{reduced stochastic
probability density function }(rs-pdf)
defined on $V_{\alpha }$ , with $V_{\alpha }$ a non-empty subset of $%
\subseteq \mathbb{R}^{k}$ and $k\geq 1;$ 2) that the fluid fields
$\left\{ Z\right\} $ will generally be considered as stochastic
functions (see Appendix).

Critical issues in turbulence theory are in principle related both
to the definition of the hidden variables $\alpha $ as well as the
determination of the corresponding rs-pdf $g.$ Both depend
necessarily on the specific assumptions concerning the source of
stochasticity$.$ Therefore, it is obvious that for both of them
the definition is non-unique. In particular, the precise
definition of $g$ may depend on the specific source (of
stochasticity). This is obvious - for example - when its origin is
either provided by boundary conditions (\textit{i.e.}, surface
forces) or respectively by volume forces: both sources, in fact,
can be prescribed in principle in arbitrary ways. It follows that,
in general, $g$ should be considered of the form
\begin{equation}
g=g(\mathbf{r,}t\mathbf{,}\alpha ), \label{non-stationar and
nonhomogeneous turbulence}
\end{equation}%
with $(\mathbf{r,}t)\in $\ $\Omega \times ${$I$}. This corresponds
to the
so-called (\emph{statistically}) \emph{nonhomogenous} and \emph{%
nonstationary turbulence}. Particular solutions are, however,
provided by the requirements that there results identically either
$g=g(t,\mathbf{\alpha
})$ (\emph{homogeneous turbulence})$,$ or $g=g(\mathbf{r,\alpha })$ (\emph{%
stationary turbulence}).

In subsequent sections (Sec.2-6) we intend to present a
mathematical formulation of turbulence in magnetofluids which
pertains specifically to the validity of both assumptions,
\textit{i.e.},
\begin{equation}
g=g(\mathbf{\alpha }).  \label{stationary and homogeneous
turbulence}
\end{equation}%
This defines the so-called \emph{homogenous and stationary
turbulence}. Possible generalizations of the theory to
nonhomogeneous and nonstationary
turbulence are discussed in Sec.7. For this purpose, all the fluid fields $%
\left\{ Z\right\} $ will be generally assumed stochastic
(\textit{i.e.}, non-deterministic), leaving unspecified the
definition of the hidden variables and - apart for the previous
requirement - the form of the rs-pdf.

\subsection{Turbulence in magnetofluids}

The phenomenon of turbulence in magnetofluids (\emph{MHD
turbulence}) is nowadays playing a major role in plasma physics
and fluid dynamics research. We refer here in particular to
incompressible magneto- and conducting fluids for which the fluid
description is appropriate (for such systems typically a
statistical description based on the microscopic dynamics is not
possible). The widespread picture of MHD-turbulence phenomena
occurring in these fluids consists of an ensemble of
finite-amplitude waves with random phase. Examples of turbulence
models of this type are well known. For example, Iroshnikov
\cite{Iroshnikov} and Kraichnan \cite{Kraichnan} independently
assumed that MHD turbulence occurs as a result of nonlinear
interactions between Alfven wave packets
\cite{Mattheus,Goldreich}. However, there is an increasing
evidence that this picture is an oversimplification. In
particular, MHD turbulence may include fluctuations whose
phase-coherence characteristics are incompatible with wave-like
properties. The latter are the so-called coherent structures, like
clumps, holes, current filaments, shocks, magnetic islands,
vortices, convective cells, zonal flows, streamers, etc.
\cite{Dupree1972,Graddock}.

In fluid turbulence the signature of the presence of coherent
structures is provided by the existence of non-Gaussian features
in the probability density. This is usually identified with the
velocity-difference probability density function, traditionally
adopted for the description of homogeneous turbulence. In the past
several statistical models have been proposed for its
determination, which include the mapping-closure method, the
test-function method and the field-theoretical approach \cite%
{Chen1989,Pope1985,Polyakov1995}. Nevertheless, despite the
progress achieved in modelling key features of the basic
phenomenology, still missing is a consistent, theory-based,
statistical description of MHD turbulence. Clearly, such
formulation - if achievable at all - should rely exclusively on a
rigorous, deductive formulation of the turbulence-modified fluid
equations following from the fluid equations.

\subsection{Goals of the investigation}

Based on a recently proposed inverse kinetic theory (IKT) for
classical and
quantum fluids (Ellero and Tessarotto, 2004-2007 \cite%
{Ellero2000,Tessarotto2004,Ellero2005,Tessarotto2006}), and in
particular its formulation for MHD equations
\cite{Tessarotto2008,Tessarotto2008a}, a phase-space statistical
model of turbulence is proposed. The present approach, which
represents a generalization of the one recently developed for
incompressible neutral fluids
\cite{Tessarotto2008z,Tessarotto2008z2},
allows to determine exactly the stochastic evolution equation of the \emph{%
local position-velocity joint probability density function} (\emph{local pdf}%
). \emph{The local pdf, rather than the velocity-increment pdf
traditionally adopted in turbulence theory, is in fact found to be
meaningful for the dynamics of the fluid.} Indeed, we intend to
prove that the stochastic fluid fields which characterize the
turbulent magnetofluid are uniquely determined by means of
suitable velocity-moments of the local pdf. Nevertheless, as
previously pointed out \cite{Tessarotto2006,Tessarotto2008z},
IKT's are intrinsically non-unique in character, in particular
because they may be also formulated in such a way to apply only to
a suitable subset of the fluid equations. In this respect two
types of phase-space approaches can in principle be developed,
namely either:

\begin{itemize}
\item a \emph{Complete IKT,} yielding the full set of stochastic fluid
equations which advance in time the stochastic fluid fields
$\left\{ Z\right\} ;$

\item a \emph{Reduced IKT}, which applies only to a suitable subset of fluid
equations, to be identified with the set fluid equations advancing
in time only $\left\{ \left\langle Z\right\rangle \right\} .$
\end{itemize}

In particular the \emph{Complete IKT} can be non-uniquely
prescribed (see THM's 1 and 2) so that:

\begin{enumerate}
\item the time-evolution of the stochastic fluid fields $\left\{ Z\right\} $
describing the state of a turbulent magnetofluid is uniquely
determined by means of a suitable phase-space classical dynamical
system (\emph{the MHD-dynamical system});

\item the MHD-dynamical system is uniquely associated to an appropriate
\emph{inverse kinetic equation} (IKE)\ which advances in time an
appropriate phase-space probability density $f$ (pdf);

\item the \emph{inverse kinetic equation} (IKE) is such that its velocity
moments coincide with the complete set of stochastic(MHD) fluid
equations.
In particular, the IKE's which advance in time the stochastic-averaged pdf $%
\left\langle f\right\rangle $ as well its stochastic fluctuation
$\delta f=f-\left\langle f\right\rangle ,$ determine respectively
the fluid equations for stochastic-averaged fluid fields $\left\{
\left\langle Z\right\rangle \right\} $ and the stochastic
fluctuations $\left\{ \delta Z\right\} =\left\{ Z-\left\langle
Z\right\rangle \right\} ;$

\item under appropriate initial and smoothness conditions,\ the strict
positivity of $f$ is assured for arbitrary strictly positive
initial pdf's;

\item under the same assumptions for $\left\langle f\right\rangle ,$
exhibits an irreversible time-evolution (see the Corollary of
THM.2)
\end{enumerate}

Instead, as previously pointed out
\cite{Tessarotto2008z,Tessarotto2008a}, the \emph{Reduced IKT}
(see THM.3) can be realized so that:

\begin{enumerate}
\item the time evolution of the stochastic averaged fluid fields is
similarly uniquely determined by means of a suitable\emph{\
}dynamical system and an appropriate inverse kinetic equation
advancing in time the stochastic-averaged pdf $\left\langle
f\right\rangle $;

\item the strict positivity of $\left\langle f\right\rangle $ is assured;

\item the inverse kinetic equation for $\left\langle f\right\rangle $ is
Markovian.
\end{enumerate}

Key feature of present theory is that - unlike several other
approaches to be found in the literature (for a review see for
example, Monin and Yaglom, 1975 \cite{Monin1975} and Pope, 2000
\cite{Pope2000}) - the relationship between fluid fields and the
pdf is \textit{exact} (and not just only approximate). In
particular, the evolution of the stochastic-averaged fluid fields
is determined via a suitable nonlinear transformation (here
generated by an appropriate dynamical system). This is achieved by
the construction of a Vlasov-type kinetic equation which advances
in time the local pdf and - as a consequence - also the stochastic
fluid fields of the magnetofluid.

\section{Motivations: IKT approach for magnetofluids}

A fundamental aspect of fluid dynamics is the construction of
phase-space approaches for realistic fluids. Indeed, phase-space
techniques are well known both in classical and quantum fluid
dynamics. In fact, generally the fluid equations represent a
mixture of hyperbolic and elliptic PDE's, which are extremely hard
to study both analytically and numerically. This has motivated in
the past efforts to replace them with other equations, possibly
simpler to solve or mathematically more elegant. In this
connection a particular viewpoint - which applies in principle
both to classical and quantum fluids - is represented by the class
of so-called \textit{inverse problems}, involving the search of a
so-called inverse kinetic theory (IKT) able to yield the complete
set of fluid equations for the fluid fields, via a suitable
\textit{correspondence principle}. As a consequence the fluid
equations are identified with appropriate moment equations
constructed in terms of the relevant kinetic equation. This raises
the interesting question whether the theory can be generalized to
arbitrary classical magnetofluids. The issue is relevant at least
for the following reasons: a) the\ proliferation of numerical
algorithms in MHD which reproduce the behavior of incompressible
fluids only in an asymptotic sense; b) the possible verification
of conjectures involving the validity of appropriate equations of
state for the fluid pressure; c) the ongoing quest for efficient
numerical solution methods for the MHD equations, with particular
reference to Lagrangian methods. Finally, another important
motivation is the possibility of achieving an exact solution to
this problem, based on inverse
kinetic theory (IKT) (see Tessarotto \textit{et al.}, 2008 \cite%
{Tessarotto2008}). In fact an aspect of fluid dynamics is
represented by the class of so-called \textit{inverse problems},
involving the search of IKT's able to yield \textit{identically} a
prescribed set of fluid equations. A few examples of IKT's which
hold both for classical and quantum fluids have
been recently investigated (Tessarotto \textit{et al.}, 2000-2007 \cite%
{Ellero2000,Tessarotto2004,Ellero2005,Tessarotto2006,
Tessarotto2007a}).

In particular, among the possible IKT's, special have IKT's which
pertain to incompressible fluids described in terms of strong
solutions of the corresponding fluid equations. In such a case
IKT's can be defined in such a way that the kinetic equation which
advances in time the kinetic
distribution function \textit{generates a suitable classical dynamical system%
}. Phase-space approaches of this type have been already achieved
for
several types of fluids (Ellero and Tessarotto, 2000-2008 \cite%
{Tessarotto2004,Ellero2005,Tessarotto2006, Tessarotto2007a}). In
particular, the same approach can be extended in principle also to
incompressible magnetofluids. Without loss of generality, in the
following we consider the case of the MHD equations describing an
incompressible, viscous, quasi-neutral, isentropic, isothermal and
resistive magnetofluid. Starting point is the assumption that
there exists a suitable phase-space classical
dynamical system, to be denoted as \emph{MHD-dynamical system},%
\begin{equation}
\mathbf{x}_{o}\rightarrow \mathbf{x}(t)=T_{t,t_{o}}\mathbf{x}_{o},
\label{classical dynamical system}
\end{equation}%
defined in such a way that\textit{\ its evolution operator }$T_{t,t_{o}}$%
\textit{\ uniquely advances in time the fluid fields}
\cite{Tessarotto2006}. The dynamical system (and the operator
$T_{t,t_{o}})$ is assumed to be generated by a suitably smooth
vector field $\mathbf{X}(\mathbf{x},t;Z),$ i.e., as a solution of
the initial-value problem
\begin{equation}
\left\{
\begin{array}{c}
\frac{d}{dt}\mathbf{x}=\mathbf{X}(\mathbf{x},t;Z), \\
\mathbf{x}(t_{o})=\mathbf{x}_{o},%
\end{array}%
\right.  \label{DYN-SYS}
\end{equation}%
where $\mathbf{X}(\mathbf{x},t;Z)$ is required to be - in an
appropriate way - functionally dependent on the set of fluid
fields $\left\{ Z\right\} $
which characterize the fluid. Here $\mathbf{x}=(\mathbf{r}_{1}\mathbf{,v}%
_{1})\in \Gamma $ and respectively $\mathbf{r}_{1}$ and
$\mathbf{v}_{1}$ denote an appropriate "state-vector" and the
corresponding "configuration" and "velocity" vectors. In
particular, it is assumed that $\mathbf{x}$ spans a phase-space of
dimension $2n$ ($\Gamma \subseteq \mathbb{R}^{2n}$), where by
assumption $n\geq 3$. Therefore, introducing the corresponding
phase-space probability density function (pdf) $f(\mathbf{x,}t;Z),$ in $%
\Gamma $ it fulfills necessarily the integral Liouville equation

\begin{equation}
J(t;Z)f(\mathbf{x}(t),t;Z)=f(\mathbf{x}_{o},t_{o},Z),
\label{Lagrangian IKE}
\end{equation}
to be viewed in the following as a \emph{Lagrangian inverse
kinetic equation}
(Lagrangian IKE), with $J(t;Z)=\left\vert \frac{\partial \mathbf{x}(t)}{%
\partial \mathbf{x}_{o}}\right\vert $ denoting the Jacobian of the map (\ref%
{classical dynamical system}). In particular, let us assume for
definiteness that (\ref{classical dynamical system}) is a
diffeomorphism at least of class $C^{(2)}(\Gamma \times I\times
I),$ with $I\subset \mathbb{R}$
denoting an appropriate finite time interval. Then, if the initial pdf $f(%
\mathbf{x}_{o},t_{o},Z)$ is suitably smooth it follows that the pdf $f(%
\mathbf{x,}t;Z)$ satisfies also the differential Liouville
equation
\begin{equation}
L(Z)f(\mathbf{x},t;Z)=0,  \label{Liouville0}
\end{equation}%
$L(Z)$ denoting the corresponding streaming operator
\begin{equation}
L(Z)\cdot \equiv \frac{\partial }{\partial t}\cdot +\frac{\partial }{%
\partial \mathbf{x}}\cdot \left\{ \mathbf{X}(\mathbf{x},t;Z)\cdot \right\} .
\end{equation}%
This equation [and the equivalent Lagrangian equation (\ref{Lagrangian IKE}%
)] will be considered in the following as as an \emph{Eulerian
inverse kinetic equation} (Eulerian IKE) for a suitable set of
fluid equations. In other words, the\ arbitrariness of the
dynamical system (\ref{classical dynamical system})
[\textit{i.e.}, of $\mathbf{X}(\mathbf{x},t;Z)$], will be
used to seek a representation of the vector field $\mathbf{X}(\mathbf{x}%
,t;Z) $ such that that suitable the velocity moments of
$f(\mathbf{x,}t;Z)$ can be identified with the relevant fluid
fields characterizing a prescribed fluid.

Let us consider, for definiteness, the case of an incompressible,
viscous, quasi-neutral, isentropic, isothermal and resistive
magnetofluid subject to the condition of isentropic flow. By
assumption, the relevant fluids
\begin{equation}
\left\{ Z\right\} \equiv \left\{ \rho =\rho _{o}>0,\mathbf{V},p\geq 0,%
\mathbf{B,}S_{T}\right\} ,  \label{fluid fields}
\end{equation}%
{\ \textit{i.e.}, respectively the mass density, the fluid
velocity, the fluid pressure, the magnetic field and the
thermodynamic entropy, are
assumed to be defined in the whole existence domain }$\overline{\Omega }%
\times I${. In particular if $\Omega $ denotes an open connected subset of }$%
R^{3},$ its closure $\overline{\Omega }$ by definition is the set
where the mass density is a constant $\rho =\rho _{o}>0${. W}e
shall assume that the fluid fields are continuous in
$\overline{\Omega }\times {I},$ {satisfy a suitable set of
\emph{deterministic MHD equations }}in the open set $\Omega \times
${$I$} and, moreover, fulfill appropriate initial and (Dirichlet)
boundary conditions respectively at $t=t_{o}$ and on the boundary
of the configuration domain $\delta \Omega ${. These requirements
are represented
respectively by the equations}%
\begin{equation}
\left\{
\begin{array}{c}
\rho =\rho _{o}, \\
\nabla \cdot \mathbf{V}=0, \\
\left. \nabla \cdot \mathbf{B}=0\right. \\
N\mathbf{V}=0, \\
\mathbf{V}\cdot N\mathbf{V}=0, \\
\left. N_{B}\mathbf{B}=0,\right. \\
\frac{\partial }{\partial t}S_{T}=0, \\
Z(\mathbf{r,}t_{o})\mathbf{=}Z_{o}(\mathbf{r}), \\
\left. Z(\mathbf{r,}t)\right\vert _{\delta \Omega }\mathbf{=}\left. Z_{w}(%
\mathbf{r,}t)\right\vert _{\delta \Omega }.%
\end{array}%
\right.  \label{(1)}
\end{equation}%
{\ }

Here{\ the notation is standard. Thus,} $N$ and $N_{B}$ are the
nonlinear
Navier-Stokes and Faraday-Neumann differential operator{\ }%
\begin{equation}
N\mathbf{V}=\frac{\partial }{\partial t}\mathbf{V}+\mathbf{V}\cdot
\nabla \mathbf{V}+\frac{1}{\rho _{o}}\left[ \nabla
p-\mathbf{f}\right] -\nu \nabla ^{2}\mathbf{V}
\end{equation}%
and
\begin{equation}
N_{B}\mathbf{B=}\frac{\partial }{\partial
t}\mathbf{B}+\mathbf{V}\cdot \nabla
\mathbf{B}-\mathbf{h-}\frac{c}{4\pi \sigma }\nabla ^{2}\mathbf{B.}
\end{equation}%
Here {the vector fields }$\mathbf{f}$ and $\mathbf{h}$, to be
denoted as
\emph{volume forces}, {read respectively }%
\begin{eqnarray}
\mathbf{f} &=&\rho _{o}\mathbf{g}+\frac{1}{4\pi }\mathbf{B\cdot
\nabla B-\nabla }\left( \frac{B^{2}}{8\pi }\right) \mathbf{,}
\label{force density}
\\
\mathbf{h} &=&-\mathbf{B\cdot \nabla V}.
\end{eqnarray}%
where the mass density $\rho _{o},$ the kinematic viscosity $\nu $
and the conductivity $\sigma $ are all real positive constants to
be considered non-stochastic. Furthermore, in the first equation,
$\rho _{o}\mathbf{g}$ denotes the gravitational force density and
the remaining terms the Lorentz force density. In addition, the
first three equation in{\ (\ref{(1)}) denote respectively the
so-called \emph{incompressibility}, \emph{isochoricity and
divergence-free }}({for }$\mathbf{B}$){\emph{\ conditions}. The
remaining ones include, instead,} {the \emph{Navier-Stokes
equation,}} the energy equation, the \emph{Faraday-Neumann
equation} for $\mathbf{B}$ and the isentropic entropy condition
(with $S_{T}$ denoting the thermodynamic entropy).{\ Finally, the
last two equations denote respectively the initial and Dirichlet
boundary conditions. }Thus, by taking the divergence of the N-S
equation, it follows the Poisson equation for the fluid pressure
$p$ which reads
\begin{equation}
\nabla ^{2}p=-\rho _{o}\nabla \cdot \left( \mathbf{V}\cdot \nabla \mathbf{V}%
\right) +\nabla \cdot \mathbf{f},  \label{5}
\end{equation}%
with $p$ to be assumed non negative and bounded in
$\overline{\Omega }\times
\overline{{I}}$. Finally, in the following and consistent with Eqs.(\ref{(1)}%
) the thermodynamic entropy $S_{T}$ will be assumed as a constant
in the whole extended configuration domain $\overline{\Omega
}\times {I}.$ \

\section{IKT for the deterministic MHD equations}

A prerequisite for the subsequent formulation of a turbulence
theory based on the phase-space approach here adopted
\cite{Tessarotto2008z} (see subsequent Sections 4-6) is the
development of a Complete IKT for the deterministic fluid
equations [see Eqs.(\ref{(1)})]. As recently pointed out in
Ref.\cite{Tessarotto2008}, this goal can be achieved by
introducing suitable generalization of the of the theory earlier
developed for the
incompressible Navier-Stokes equations \cite%
{Tessarotto2004,Ellero2005,Tessarotto2006,Tessarotto2007a}. For
definiteness, let us introduce the notations
$\mathbf{r}_{1}\mathbf{=}\left( \mathbf{r,s}\right) ,$
$\mathbf{v}_{1}\mathbf{=}\left( \mathbf{v,y}\right) $
and $\mathbf{X}=\left\{ \mathbf{v}_{1},\mathbf{F}_{1}(\mathbf{x}%
,t;f,Z)\right\} ,$ where the vectors $\mathbf{r}$ and
$\mathbf{v}_{1}$ span, respectively, the whole configuration
domain of the fluid ($\overline{\Omega
}$) and the 3-dimensional velocity space\textbf{\ }($\mathbf{%
%TCIMACRO{\U{211d} }%
%BeginExpansion
\mathbb{R}
%EndExpansion
}^{3}$). Moreover $\mathbf{s\in }\overline{\Omega },$
$\mathbf{y}\in
%TCIMACRO{\U{211d} }%
%BeginExpansion
\mathbb{R}
%EndExpansion
^{3}$ are two additional real vector variables, with $\mathbf{s}$
denoting in particular an ignorable configuration-space vector
[both for the fluid fields and the kinetic distribution function
$f(\mathbf{x,}t;Z)$]. The streaming operator $L(Z)$ in this case
reads
\begin{equation}
L(Z)\equiv \frac{\partial }{\partial t}+\mathbf{v\cdot }\frac{\partial }{%
\partial \mathbf{r}}+\frac{\partial }{\partial \mathbf{v}}\cdot \left\{
\mathbf{F}(\mathbf{x},t;f,Z)\right\} +\frac{\partial }{\partial \mathbf{y}}%
\cdot \left\{ \mathbf{Y}(\mathbf{x},t;Z)\right\} ,
\label{Streaming operator}
\end{equation}%
where $\left\{ Z\right\} $ are the fluid fields and the vector
field
\begin{equation}
\mathbf{F}_{1}(\mathbf{x},t;f,Z)=\left\{ \mathbf{F}(\mathbf{x},t;f,Z),%
\mathbf{Y}(\mathbf{x},t;f,Z)\right\}  \label{mean-field force}
\end{equation}%
can be interpreted as an effective \emph{mean field force} acting
on a particle state $\mathbf{x}=(\mathbf{r,s},\mathbf{v,y}).$ In
the following we intend to prove that, at least in a suitable
finite time-interval $I$, the fluid fields
$Z(\mathbf{r,}t\mathbb{)}$ can be uniquely identified with the
velocity moments $\int\limits_{%
%TCIMACRO{\U{211d} }%
%BeginExpansion
\mathbb{R}
%EndExpansion
^{n}}d\mathbf{v}_{1}G(\mathbf{x,}t)f(\mathbf{x,}t;Z),$ where $f(\mathbf{x,}%
t;Z)$ is properly defined kinetic distribution function and
$G(\mathbf{x,}t)$
appropriate weight functions. More precisely, there results:%
\begin{equation}
\left\{
\begin{array}{c}
1=\int d\mathbf{v}d\mathbf{y}f(\mathbf{x},t;Z), \\
\mathbf{V}(\mathbf{r,}t)=\int d\mathbf{v}d\mathbf{\mathbf{y}v}f(\mathbf{x}%
,t;Z), \\
p_{1}(\mathbf{r,}t)=\rho _{o}\int d\mathbf{v}d\mathbf{y}\frac{1}{3}u^{2}f(%
\mathbf{x},t;Z), \\
\mathbf{B}(\mathbf{r,}t)=\int d\mathbf{v}d\mathbf{\mathbf{y}y}f(\mathbf{x}%
,t;Z), \\
S_{T}(t)=S(f(t)),%
\end{array}%
\right.  \label{moments-0}
\end{equation}%
(\emph{correspondence principle}). In particular $\mathbf{V},p$ and\ $%
\mathbf{B}$ are identified respectively with the moments $G(\mathbf{x,}t)=%
\mathbf{v},\rho _{o}u^{2}/3,$ $\mathbf{y,}$ where
$\mathbf{u=v-V}.$ In addition, if the same distribution function
$f(t)\equiv f(\mathbf{x,}t;Z)$ is strictly positive in the whole
set $\Gamma \times I,$ $S(f(t))$ is the statistical
Boltzmann-Shannon entropy functional
\begin{equation}
S(f(t))=-\int\limits_{\Gamma }d\mathbf{x}f(\mathbf{x,}t;Z)\ln f(\mathbf{x,}%
t;Z).  \label{Boltzmann-Shannon}
\end{equation}%
Consistent with Eqs.(\ref{(1)}) we shall impose that $S(f(t))$
exists for all $t\in I,$ and that there results identically
($\forall t\in I$) the conservation law
\begin{equation}
S(f(t))=const,
\end{equation}%
(\emph{constant H-theorem}). To reach the proof, let us first show
that, by
suitable definition of the "force" fields $\mathbf{F}(\mathbf{x},t)$ and $%
\mathbf{Y}(\mathbf{x},t),$ a particular solution (which defines a
kinetic equilibrium \cite{Kinetic equilibrium}) of the IKE
(\ref{Liouville0}) is delivered by the Maxwellian distribution:
\begin{equation}
f_{M}(\mathbf{x},t;Z)=\frac{\rho }{\pi
^{2}v_{th,p}^{3}v_{th,T}}\exp \left\{ -X^{2}-Y^{2}\right\} .
\label{Maxwellian0}
\end{equation}%
Here we denote
\begin{equation}
X^{2}=\frac{u^{2}}{vth_{p}^{2}},  \label{pos-1}
\end{equation}%
\begin{equation}
Y^{2}=\frac{w^{2}}{v_{th,T}^{2}},  \label{pos-2}
\end{equation}%
where $\mathbf{u=v-V,}$ $\mathbf{w=y-B,}$
$v_{th}^{2}=2\widehat{p}_{1}/\rho _{o}$ and $v_{th,T}^{2}$ is an
'\textit{a priori}' arbitrary strictly positive constant.
Furthermore
\begin{equation}
p_{1}(\mathbf{r},t)=p_{0}(t)+p(\mathbf{r},t)  \label{kinetic
pressure-1}
\end{equation}%
is the kinetic pressure. In these definitions, $p_{0}(t)$ (to be
denoted as \emph{pseudo-pressure}) is an arbitrary strictly
positive and suitably smooth function defined in $I,$ while the
mass $m>0$ is an arbitrary real constant. The following theorem
can immediately be proven:

\textbf{Theorem 1 - Complete IKT for the deterministic MHD equations } \emph{%
Let us assume that in the existence domain }$\Omega \times I$
\emph{the fluid fields (\ref{fluid fields}) are strong solutions
of the MHD fluid equations (\ref{(1)}). Furthermore, let us
identify} \emph{the vector fields }$\mathbf{F(x},t;f,Z)$\emph{\
and }$\mathbf{Y}(\mathbf{x},t;f,Z)$\emph{\
with }%
\begin{equation}
\mathbf{F}(\mathbf{x},t;f_{M},Z)=\mathbf{F}_{0}+\mathbf{F}_{1},
\label{G-1}
\end{equation}

\begin{equation}
\mathbf{Y}(\mathbf{\mathbf{x},}t;f_{M},Z)\mathbf{=u\cdot \nabla B}-\mathbf{h-%
}\frac{c}{4\pi \sigma }\nabla ^{2}\mathbf{B+w}\nabla \cdot
\mathbf{B,} \label{G-2}
\end{equation}

\emph{where }$\mathbf{F}_{0}$ \emph{and}\textbf{\
}$\mathbf{F}_{1}$\emph{\
read respectively }%
\begin{equation}
\mathbf{F}_{0}(\mathbf{x,}t;f_{M},Z)=\frac{1}{\rho _{o}}\mathbf{f}+\frac{1}{2%
}\mathbf{u}\cdot \nabla \mathbf{V+}\frac{1}{2}\mathbb{\nabla
}\mathbf{V\cdot u+}\nu \nabla ^{2}\mathbf{V,}  \label{G-3}
\end{equation}%
\begin{equation}
\mathbf{F}_{1}(\mathbf{x,}t;f_{M},Z)=\frac{\mathbf{u}}{2p_{1}}\frac{D}{Dt}%
p_{1}+\frac{v_{th}^{2}}{2}\nabla \ln p_{1}\left\{ \frac{\mathbf{u}^{2}}{%
v_{th}^{2}}-\frac{3}{2}\right\} ,  \label{G-4}
\end{equation}%
\emph{while }%
\begin{equation}
\frac{D}{Dt}\equiv \frac{\partial }{\partial
t}+\mathbf{V}\boldsymbol{\cdot \nabla }
\end{equation}%
\emph{and }$\frac{D}{Dt}p_{1}$ \emph{is uniquely prescribed
consistent with
the energy equation, namely }%
\begin{equation}
\frac{D}{Dt}p_{1}=\frac{\partial }{\partial t}p_{1}-\rho _{o}\left[ \frac{%
\partial }{\partial t}V^{2}/2+\mathbf{V}\cdot \nabla V^{2}/2-\frac{1}{\rho
_{o}}\mathbf{V\cdot f}-\nu \mathbf{V\cdot }\nabla
^{2}\mathbf{V}\right] .
\end{equation}%
\emph{Finally, let us assume that the initial kinetic probability density }$%
f(t_{o})\equiv f(\mathbf{x,}t_{o},Z)$\emph{\ belongs to the
functional class}
$\left\{ f(t_{o})\right\} ,$\emph{\ defined so that }$f(t_{o})$ \emph{%
fulfils at time }$t=t_{o}$ \emph{the moment equations
(\ref{moments-0}),
while the initial fluid fields }$Z(\mathbf{r,}t_{o})\mathbf{=}Z_{o}(\mathbf{r%
})$\emph{\ are suitably prescribed}.

\emph{It follows that:}

\emph{1) the local Maxwellian distribution (\ref{Maxwellian0}) is
a particular solution of IKE [Eq.(\ref{Liouville0})]};

\emph{2) this is a solution of the same equation if an only if the
fluid
fields }$\left\{ Z\right\} $\emph{\ satisfy the fluid equations {(\ref{(1)})}%
;}

\emph{3) the velocity-moment equations obtained by taking the
weighted
velocity integrals of Eq.(\ref{Liouville0}) with the weights }%
\begin{equation}
G(x,t)=\rho _{o},\mathbf{v},\rho _{o}u^{2}/3,\mathbf{y,}w^{2}/3
\label{weights}
\end{equation}%
\emph{\ deliver identically the same fluid equations
{(\ref{(1)});}}

\emph{4) the initial kinetic probability density }$f(t_{o})\equiv f(\mathbf{%
x,}t_{o},Z)$\emph{\ is assumed to satisfy the principle of entropy
maximization (PEM) \cite{Jaynes1957}, which requires that }%
\begin{equation}
\delta S(f(t_{o}))=0;  \label{PEM}
\end{equation}

\emph{5) }$\forall t,t_{o}\in I,$ \emph{the statistical entropy
}$S(f(t))$
\emph{satisfies the constraint}%
\begin{equation}
S(f(t_{o}))=S(f(t))  \label{constant-H theorem}
\end{equation}%
(\emph{constant-H theorem}).

\noindent \emph{PROOF}

First, we notice that if the fluid equations (\ref{(1)}) are
satisfied identically in $\Omega \times I,$ the proof that
(\ref{Maxwellian0}) is a particular solution of the IKE
[Eq.(\ref{Liouville0})] follows by direct
differentiation. The converse implication, \textit{i.e.}, the proof that if (%
\ref{Maxwellian0}) is a solution of Eq.(\ref{Liouville0}) then the
fluid equations {(\ref{(1)}) }are satisfied identically in $\Omega
\times I,$ follows by evaluating in particular the
($\mathbf{v}_{1}-$)velocity-space moments of Eq.(\ref{Liouville0})
for the weights (\ref{weights}) $G=\rho _{o}\left(
\mathbf{v-V}\right) ^{2}/3$ and $w^{2}/3.$ These moments deliver
respectively the two isochoricity conditions%
\begin{eqnarray}
\nabla \cdot \mathbf{V} &=&0, \\
\nabla \cdot \mathbf{B} &=&0.
\end{eqnarray}

The proof of Proposition 3) follows by noting that the extremal
solution of the variational principle (\ref{PEM}) coincides
necessarily with the
Maxwellian distribution $f_{M}(\mathbf{x},t_{o};Z)$ [see Eq.(\ref%
{Maxwellian0})]. Finally, to satisfy Proposition 4) it is
sufficient to notice that the pseudo-pressure $p_{o}(t)$ [which
enters the definition of the kinetic pressure (\ref{kinetic
pressure-1})] can always we defined in such a way to satisfy, at
least in a finite time interval $I,$ the constraint
(\ref{constant-H theorem}). Q.E.D.

In analogy to the case of isothermal fluids the present theorem
can be generalized in a straightforward way to a suitably smooth
non-Maxwellian initial distribution function \cite{Ellero2005}.

\section{Stochastic MHD fluid equations}

Let us now construct the stochastic MHD fluid equations
appropriate in the case of homogeneous and stationary turbulence.
In principle this requires the adoption of an appropriate
\emph{turbulence} (or stochastic)\emph{\ model.} This concerns the
introduction of suitable hypotheses, \textit{i.e.},

\begin{enumerate}
\item \emph{the definition of an appropriate set of hidden variables }$%
\mathbf{\alpha }$;

\item \emph{the definition of the associated rs-pdf }$g;$

\item \emph{the assumption that at least some of the fluid fields, together
with their initial and boundary values on }$\delta \Omega
,$\emph{\ and/or the volume forces }$f$\emph{\ and }$h$\emph{\
depend in a suitable way from the hidden variables
}$\mathbf{\alpha };$\emph{\ }

\item \emph{the prescription of the form of the stochastic fluid equations.}
\end{enumerate}

Its should be stressed that in principle the definition of the $\mathbf{%
\alpha }^{\prime }s$ - and consequently\emph{\ }of\emph{\ }$g$ and
of the dependencies on $\mathbf{\alpha }$ to be suitably
prescribed in the fluid equations - remains completely arbitrary.
Regarding item 4), in particular, it is obvious that '\textit{a
priori} ' also the stochastic fluid equations might differ from
the deterministic equations (\ref{(1)}). Thus, for example, the
condition of incompressibly of the fluid $\rho =\rho _{o}$ might
be violated in a turbulent fluid due to the possible presence of
stochastic mass density fluctuations (see Sec.7).

For definiteness, let us assume that the fluid fields are
stochastic, i.e., they depend smoothly on the hidden variables
$\mathbf{\alpha ,}$ while the stochastic fluid equations retain
the same functional form of the corresponding deterministic
equations (\ref{(1)}). This implies that the fluid remains, in
particular, incompressible, viscous, quasi-neutral, isentropic,
isothermal and resistive magnetofluid, which - in particular -
implies necessarily $\delta \rho \mathbb{\equiv }0$ and $\delta
S_{T}\equiv 0.$ Hence the meaningful fluid fields are only
represented by the subset
\begin{equation}
\left\{ Z(\mathbf{\alpha }\mathbb{)}\right\} \mathbb{\equiv }\left\{ \mathbf{%
V}(\mathbf{r,}t,\mathbf{\alpha }\mathbb{)},p(\mathbf{r,}t,\mathbf{\alpha }%
\mathbb{)}\geq 0,\mathbf{B}(\mathbf{r,}t,\mathbf{\alpha
}\mathbb{)}\right\} .
\end{equation}%
Together with their initial and boundary values on $\delta \Omega
$\
(defined respectively by the vector fields $Z_{o}(\mathbf{r},\mathbf{\alpha }%
\mathbb{)}$ and $\left. Z(\mathbf{r,}t,\mathbf{\alpha
}\mathbb{)}\right\vert
_{\delta \Omega }\equiv \left. Z_{w}(\mathbf{r,}t,\mathbf{\alpha }\mathbb{)}%
\right\vert _{\delta \Omega })${, and }the volume forces $\mathbf{f}(\mathbf{%
r,}t,\mathbf{\alpha }\mathbb{)},$ $\mathbf{h}(\mathbf{r,}t,\mathbf{\alpha }%
\mathbb{)},$ we shall require that they all admit the
\emph{stochastic decompositions:}
\begin{eqnarray}
Z(\mathbf{r,}t;\mathbf{\alpha }\mathbb{)} &\mathbb{=}&\left\langle Z(\mathbf{%
r,}t,\mathbf{\alpha }\mathbb{)}\right\rangle +\delta Z(\mathbf{r,}t,\mathbf{%
\alpha }\mathbb{)}, \\
Z_{o}(\mathbf{r;\alpha }\mathbb{)} &\mathbb{=}&\left\langle Z_{o}(\mathbf{r\,%
},\mathbf{\alpha }\mathbb{)}\right\rangle +\delta Z_{o}(\mathbf{r},\mathbf{%
\alpha }\mathbb{)}, \\
Z_{w}(\mathbf{r,}t,\mathbf{\alpha }\mathbb{)} &\mathbb{=}&\left\langle Z_{w}(%
\mathbf{r,}t,\mathbf{\alpha }\mathbb{)}\right\rangle +\delta Z_{w}(\mathbf{r,%
}t,\mathbf{\alpha }\mathbb{)}, \\
\mathbf{f}(\mathbf{r,}t,\mathbf{\alpha }\mathbb{)}
&\mathbb{=}&\left\langle \mathbf{f}(\mathbf{r,}t,\mathbf{\alpha
}\mathbb{)}\right\rangle +\delta
\mathbf{f}(\mathbf{r,}t,\mathbf{\alpha }\mathbb{)}.
\end{eqnarray}%
Here the stochastic-averaging operator $\left\langle
{}\right\rangle $ is defined according to Eq.(\ref{stochastic
agerage}), while $\left\langle Z\right\rangle ,$ $\left\langle
Z_{o}\right\rangle ,$ $\left\langle Z_{w}\right\rangle ,$
$\left\langle \mathbf{f}\right\rangle $ and $\delta Z,$ $\delta
Z_{o},$ $\delta Z_{w}$ and $\delta \mathbf{f}$ are denoted
respectively as \emph{stochastic averages} and \emph{stochastic fluctuations}%
.

In the remainder the precise definition of the operator
$\left\langle {}\right\rangle $\ [\textit{i.e.}, the specification
of $g(\mathbf{\alpha })$ and of the hidden variables
$\mathbf{\alpha }$] is not needed. The only
requirement to be introduced is that - consistent with the assumption (\ref%
{stationary and homogeneous turbulence}) - it commutes with all
the
differential operators appearing in the MHD equations, in particular, $\frac{%
\partial }{\partial t},$ $\nabla $ and $\nabla ^{2}$. In such a case the
fluids equations for the averaged fields $\left\{ \left\langle Z(\mathbf{%
\alpha }\mathbb{)}\right\rangle \right\} $, {are immediately found to be}%
\begin{equation}
\left\{
\begin{array}{c}
\left\langle \rho \right\rangle =\rho _{o}, \\
\nabla \cdot \left\langle \mathbf{V}\right\rangle =0, \\
\left. \nabla \cdot \left\langle \mathbf{B}\right\rangle =0,\right. \\
\left\langle N\right\rangle \left\langle \mathbf{V}\right\rangle
+\left\langle \delta N\delta \mathbf{V}\right\rangle =0, \\
\left\langle \mathbf{V}\right\rangle \cdot \left[ \left\langle
N\right\rangle \left\langle \mathbf{V}\right\rangle +\left\langle
\delta
N\delta \mathbf{V}\right\rangle \right] =0, \\
\left. \left\langle N_{B}\right\rangle \left\langle
\mathbf{B}\right\rangle
+\left\langle \delta N_{B}\delta \mathbf{B}\right\rangle =0,\right. \\
\frac{\partial }{\partial t}\left\langle S_{T}\right\rangle =0, \\
\left\langle Z(\mathbf{r,}t_{o};\mathbf{\alpha })\right\rangle \mathbf{=}%
\left\langle Z_{o}(\mathbf{r};\mathbf{\alpha })\right\rangle , \\
\left. \left\langle Z(\mathbf{r,}t;\mathbf{\alpha })\right\rangle
\right\vert _{\delta \Omega }\mathbf{=}\left. \left\langle Z_{w}(\mathbf{r,}%
t;\mathbf{\alpha })\right\rangle \right\vert _{\delta \Omega }\mathbf{,}%
\end{array}%
\right.  \label{8}
\end{equation}%
{\ which are }denoted as the \emph{stochastic-averaged MHD equations }for%
\emph{\ }$\left\{ \left\langle Z\right\rangle \right\} $.
Analogous
equations hold for the stochastic fluctuations\emph{\ }$\left\{ \delta Z(%
\mathbf{\alpha }\mathbb{)}\right\} :$%
\begin{equation}
\left\{
\begin{array}{c}
\nabla \cdot \delta \mathbf{V}=0, \\
\left. \nabla \cdot \delta \mathbf{B}=0,\right. \\
\left\langle N\right\rangle \delta \mathbf{V}+\delta N\delta \mathbf{V-}%
\left\langle \delta N\delta \mathbf{V}\right\rangle =0, \\
\left\langle \mathbf{V}\right\rangle \cdot \left[ \left\langle
N\right\rangle \delta \mathbf{V}+\delta N\delta
\mathbf{V-}\left\langle
\delta N\delta \mathbf{V}\right\rangle \right] =0, \\
\left. \left\langle N_{B}\right\rangle \delta \mathbf{B}+\delta
N_{B}\delta \mathbf{B-}\left\langle \delta N_{B}\delta
\mathbf{B}\right\rangle =0,\right.
\\
\delta Z(\mathbf{r,}t_{o};\mathbf{\alpha })\mathbf{=\delta }Z_{o}(\mathbf{r};%
\mathbf{\alpha }), \\
\left. \delta Z(\mathbf{r,}t;\mathbf{\alpha })\right\vert _{\delta \Omega }%
\mathbf{=}\left. \delta Z_{w}(\mathbf{r,}t;\mathbf{\alpha
})\right\vert
_{\delta \Omega }%
\end{array}%
\right.  \label{8b}
\end{equation}%
(\emph{stochastic-fluctuating MHD equations }). Here the notation
is standard \cite{Tessarotto2008,Tessarotto2008a}. Finally,
$\left\langle N\right\rangle ,$ $\delta N$ and respectively
$\left\langle N_{B}\right\rangle ,$ $\delta N_{B}$ are the
nonlinear operators
\begin{equation}
\left\langle N\right\rangle \left\langle \mathbf{V}\right\rangle =\frac{%
\partial }{\partial t}\left\langle \mathbf{V}\right\rangle +\left\langle
\mathbf{V}\right\rangle \cdot \nabla \left\langle \mathbf{V}\right\rangle +%
\frac{1}{\rho _{o}}\left[ \nabla \left\langle p\right\rangle
-\left\langle
\mathbf{f}\right\rangle \right] -\nu \nabla ^{2}\left\langle \mathbf{V}%
\right\rangle ,
\end{equation}%
\textbf{\ \ }%
\begin{equation}
\delta N\delta \mathbf{V}=\delta \mathbf{V}\cdot \nabla \delta \mathbf{V}+%
\frac{1}{\rho _{o}}\left[ \nabla \delta p-\delta \mathbf{f}\right]
-\nu \nabla ^{2}\delta \mathbf{V,}
\end{equation}%
and
\begin{equation}
\left\langle N_{B}\right\rangle \left\langle \mathbf{B}\right\rangle =\frac{%
\partial }{\partial t}\left\langle \mathbf{B}\right\rangle +\left\langle
\mathbf{V}\right\rangle \cdot \nabla \left\langle
\mathbf{B}\right\rangle -\left\langle \mathbf{h}\right\rangle
\mathbf{-}\frac{c}{4\pi \sigma }\nabla ^{2}\left\langle
\mathbf{B}\right\rangle ,
\end{equation}%
\begin{equation}
\delta N_{B}\delta \mathbf{B}=\delta \mathbf{V}\cdot \nabla \delta \mathbf{B}%
-\delta \mathbf{h}-\frac{c}{4\pi \sigma }\nabla ^{2}\delta
\mathbf{B}.
\end{equation}

\section{Complete IKT formulation for the stochastic MHD equations}

As indicated above (see also related discussion in Ref. \cite%
{Tessarotto2008z}) in principle the problem of the formulation of
a stochastic IKT can be posed either for the full set of fluid
equations or only for the stochastic-averaged equations (\ref{8})
(see also Sec.6).

In the first case let us assume, for greater generality, that the
pdf can depend on the hidden variables both implicitly, via the
fluid fields, and
possibly also explicitly, i.e., it is stochastic local pdf of the form $f(%
\mathbf{x,}t;Z(\mathbf{\alpha }),\mathbf{\alpha }\mathbb{)}$. The
formulation of the IKT in this case follows directly from THM.1 by
imposing that $f(\mathbf{x,}t;Z(\mathbf{\alpha }),\mathbf{\alpha
}\mathbb{)}$ satisfies the stochastic Eulerian IKE
\begin{equation}
L(Z(\mathbf{\alpha }))f(\mathbf{x,}t;Z(\mathbf{\alpha }),\mathbf{\alpha }%
\mathbb{)}=0.  \label{stochastic IKE}
\end{equation}%
Here, the streaming operator $L(Z(\mathbf{\alpha }))$ is again
defined by
Eq.(\ref{Streaming operator}), while by assumption the vector fields $%
\mathbf{F}$ and $\mathbf{Y}$ depend on $\mathbf{\alpha }$ only
implicitly via the fluid fields $\left\{ Z(\mathbf{\alpha
})\right\} $. \ Next, let us introduce for the pdf and the
streaming operator the stochastic decompositions
\begin{eqnarray}
&&\left. f(t;\mathbf{\alpha }\mathbb{)\equiv }f(\mathbf{x,}t;Z(\mathbf{%
\alpha }),\mathbf{\alpha }\mathbb{)}\mathbb{=}\left\langle f(\mathbf{x,}t;Z(%
\mathbf{\alpha }),\mathbf{\alpha }\mathbb{)}\right\rangle +\delta f(\mathbf{%
x,}t;Z(\mathbf{\alpha }),\mathbf{\alpha }\mathbb{)},\right.
\label{STOC-1}
\\
&&\left. L(Z(\mathbf{\alpha }))=\left\langle L(Z(\mathbf{\alpha }%
))\right\rangle +\delta L(Z(\mathbf{\alpha })),\right.
\label{STOC-2}
\end{eqnarray}%
where $\left\langle {}\right\rangle $ is again the
stochastic-averaging operator (\ref{stochastic agerage}).
Requiring that $\left\langle {}\right\rangle $ commutes also with
the differential operators $\partial
/\partial \mathbf{v}$ and $\partial /\partial \mathbf{y,}$ Eq.(\ref%
{stochastic averaging operator}) yields the two stochastic kinetic
equations advancing in time respectively $\left\langle
f\right\rangle \equiv
\left\langle f(\mathbf{x,}t;Z(\mathbf{\alpha }),\mathbf{\alpha }\mathbb{)}%
\right\rangle $ and $\delta f\equiv \delta f(\mathbf{x,}t;Z(\mathbf{\alpha }%
),\mathbf{\alpha }\mathbb{)},\mathbb{\ }$\textit{i.e.}
\begin{equation}
\left\langle L(Z)\right\rangle \left\langle f\right\rangle
+\left\langle \delta L(Z)\delta f\right\rangle =0,  \label{IKE-1}
\end{equation}%
\begin{equation}
\left\langle L(Z)\right\rangle \delta f+\delta L(Z)\delta
f-\left\langle \delta L(Z)\delta f\right\rangle =0.  \label{IKE-2}
\end{equation}%
Thanks to THM.1, Eq.(\ref{stochastic IKE}) [or the equivalent Eqs.(\ref%
{IKE-1}) and (\ref{IKE-2})] provides, in principle, the exact (and
unique)
evolution of the stochastic pdf $f(\mathbf{x,}t;Z(\mathbf{\alpha }),\mathbf{%
\alpha }\mathbb{)}.\mathbb{\ }$ Furthermore, thanks to the
correspondence
principle [given by Eqs.(\ref{moments-0})], also of stochastic fluid fields $%
\left\{ Z\right\} $ are uniquely determined
\cite{Tessarotto2008z}.

In this Section we intend to formulate a \emph{Complete IKT,}
yielding the full set of stochastic fluid equations advancing in
time the stochastic fluid fields $\left\{ Z\right\} ,$ here
represented by the stochastic MHD problem defined above. \ The
basic result has the flavor of:

\textbf{Theorem 2 - Complete IKT for the stochastic MHD equations } \emph{%
Requiring that the fluid fields} $\left\{ Z(\mathbf{\alpha
})\right\} \mathbb{\equiv }\left\{ \mathbf{V},p\geq
0,\mathbf{B}\right\} $ \emph{are stochastic [in the sense the of
the definition given above], and the
correspondence principle}%
\begin{equation}
\left\{
\begin{array}{c}
1=\int d\mathbf{v}d\mathbf{y}f(\mathbf{x,}t;Z(\mathbf{\alpha }),\mathbf{%
\alpha }\mathbb{)}, \\
\mathbf{V}(\mathbf{r,}t;\mathbf{\alpha })=\int d\mathbf{v}d\mathbf{\mathbf{y}%
v}f(\mathbf{x,}t;Z(\mathbf{\alpha }),\mathbf{\alpha }\mathbb{)}, \\
p_{1}(\mathbf{r,}t)=\rho _{o}\int d\mathbf{v}d\mathbf{y}\frac{1}{3}\mathbf{u}%
^{2}f(\mathbf{x,}t;Z(\mathbf{\alpha }),\mathbf{\alpha }\mathbb{)}, \\
\mathbf{B}(\mathbf{r,}t;\mathbf{\alpha })=\int d\mathbf{v}d\mathbf{\mathbf{y}%
y}(\mathbf{x,}t;Z(\mathbf{\alpha }),\mathbf{\alpha }\mathbb{)}, \\
S_{T}(t;\mathbf{\alpha })=S(f(t;\mathbf{\alpha })),%
\end{array}%
\right.  \label{corr-principle}
\end{equation}%
\emph{holds identically in }$\overline{\Gamma }\times I$\emph{\ (where }$%
\overline{\Gamma }$\emph{\ is the closure of }$\Gamma =\Omega \times \mathbb{%
R}^{3}$\emph{)} \emph{then provided the mean-field force }
\begin{equation}
\mathbf{F}_{1}(\mathbf{x},t;f,Z(\mathbf{\alpha }),\mathbf{\alpha
})=\left\{
\mathbf{F}(\mathbf{x},t;f,Z(\mathbf{\alpha }),\mathbf{\alpha }),\mathbf{Y}(%
\mathbf{x},t;f,Z(\mathbf{\alpha }),\mathbf{\alpha })\right\}
\label{mean-field-force- stochastic}
\end{equation}%
\emph{\ is still defined by Eqs.(\ref{G-1})- (\ref{G-3}) (with
fluid fields to be considered stochastic functions) it follows
that:}

\emph{1) the local distribution }$f_{M}(\mathbf{x},t;Z(\mathbf{\alpha })),$%
\emph{\ \textit{i.e.}, the Maxwellian distribution
(\ref{Maxwellian0}) with stochastic fluid fields [or,equivalent,
the contributions }$\left\langle
f_{M}(\mathbf{x},t;Z(\mathbf{\alpha }))\right\rangle $ \emph{and}
$\delta f_{M}(\mathbf{x},t;Z(\mathbf{\alpha }))$\emph{] is a
particular solution of
the inverse kinetic equation (\ref{stochastic IKE}) [respectively of Eqs.(%
\ref{IKE-1})-(\ref{IKE-2})] if only if the fluid fields }$\left\{
\left\langle Z(\mathbf{\alpha })\right\rangle \right\} $\emph{\ and} $%
\left\{ \delta Z(\mathbf{\alpha })\right\} $ \emph{satisfy
respectively the stochastic fluid equations (\ref{8}) and
(\ref{8b});}

\emph{2)} $f_{M}(\mathbf{x},t;Z(\mathbf{\alpha }))$
\emph{maximizes the Boltzmann-Shannon entropy }$S(f(t;\alpha
))$\emph{\ in the functional class}
$\left\{ f(\mathbf{x},t;\left\langle Z(\mathbf{\alpha )}\right\rangle ,%
\mathbf{\alpha })\right\} $ \emph{defined so that the pdf
satisfies solely the constraints (\ref{corr-principle}), where the
stochastic-averaged fluid fields} $\left\{ Z(\mathbf{\alpha
})\right\} $ \emph{are considered prescribed;}

\emph{3) the velocity-moment equations obtained by taking the
weighted
velocity integrals of Eq.(\ref{stochastic IKE}) [or, equivalent, Eqs.(\ref%
{IKE-1})-(\ref{IKE-1})] with the weights (\ref{weights})\ deliver
identically the same stochastic fluid equations (\ref{8}) and
(\ref{8b}).}

\noindent \emph{PROOF}

The proof is an immediate consequence of THM.1 and follows by
invoking the stochastic decompositions (\ref{STOC-1}) and
(\ref{STOC-2}), together with
the assumption (\ref{stationary and homogeneous turbulence}) for the rs-pdf $%
g,$ which assures the equivalence of the two stochastic kinetic equations (%
\ref{IKE-1})-(\ref{IKE-1}) with Eq.(\ref{Liouville0})$.$ In
particular, it
is immediate by direct evaluation to verify that stochastic fluid equations (%
\ref{8}) and (\ref{8b}) are recovered from
Eqs.(\ref{IKE-1})-(\ref{IKE-1}) by taking their moments
(\ref{weights}). Q.E.D.

Main consequences of THM.2 are that:

\begin{enumerate}
\item the inverse kinetic equation, either represented in the Eulerian or
Lagrangian forms. In particular, the Eulerian equation is provided by Eq.(%
\ref{stochastic IKE})\ [or by the equivalent Eqs.(\ref{IKE-1}) and (\ref%
{IKE-2})], while the Lagrangian equation, in view of Eq.(\ref{Lagrangian IKE}%
), is again of the form (\ref{Lagrangian IKE}), namely
\begin{equation}
J(t;Z(\mathbf{\alpha }),\mathbf{\alpha })f(\mathbf{x}(t,\mathbf{\alpha }%
),t;Z(\mathbf{\alpha }),\mathbf{\alpha })=f(\mathbf{x}_{o},t_{o},Z(\mathbf{%
\alpha }),\mathbf{\alpha })  \label{STOCH LAGRANGIAN IKE}
\end{equation}

where $\mathbf{x}_{o}\rightarrow \mathbf{x}(t,\mathbf{\alpha })$
is the \emph{stochastic MHD-dynamical system} determined by the
(stochastic)
phase-space Lagrangian equation (\ref{DYN-SYS}) with vector field $\mathbf{X}%
(\mathbf{x},t;Z(\mathbf{\alpha })),$ $J(t;Z(\mathbf{\alpha
}),\mathbf{\alpha
})=\left\vert \frac{\partial \mathbf{x}(t)}{\partial \mathbf{x}_{o}}%
\right\vert $ denoting the corresponding Jacobian.

\item the two phase-space representations, provided respectively by the
Eulerian and Lagrangian IKE's, are equivalent. In fact, by
construction, both equations satisfy identically the stochastic
MHD equations, including the appropriate initial and boundary
conditions imposed on the fluid fields;

\item the time-evolution of the stochastic pdf $f\mathbb{\ }$is, in turn,
uniquely determined by the stochastic dynamical system defined
above [see, in fact, the Lagrangian inverse kinetic equation
Eq.(\ref{STOCH LAGRANGIAN IKE})].

\item The stochastic-averaged pdf $\left\langle f\right\rangle $ can be
proven to be strictly positive. This result is important to assure
both that $\left\langle f\right\rangle $ is truly a probability
density and that its time evolution is (possibly) irreversible.
This result can be shown to be a consequence of THM.1.
\end{enumerate}

\section{Markovian Reduced IKT for the stochastic MHD equations}

In the traditional approach to turbulence theory main emphasis is
related to the construction of the statistical equation advancing
in time the stochastic-averaged pdf, rather than on the complete
pdf. An example is provided by stochastic models - based on
Markovian Fokker-Planck models of small-scale fluid turbulence
recently investigated in the literature by several authors
(including: Naert \textit{et al.}, 1997 \cite{Naert1997};
Friedrich and Peinke \textit{et al.}, 1999 \cite{Friedrich1999};
Luck
\textit{et al.}, 1999 \cite{Luck1999}; Cleve \textit{et al.}, 2000 \cite%
{Cleve2000}; Ragwitz and Kantz, 2001 \cite{Ragwitz2001}; Renner
\textit{et
al., }2001, 2002 \cite{Renner2001,Renner2002}; Hosokawa, 2002 \cite%
{Hosokawa2002}).

An interesting issue is whether the possible Markovian character
of the statistical equation advancing in time the
stochastic-averaged pdf, in the present case to be identified with
the local pdf $\left\langle f\right\rangle ,$ can be established
based on an inverse kinetic theory approach.

As previously pointed out \cite{Tessarotto2008z,Tessarotto2008z2},
the stochastic IKE provided by Eq.(\ref{stochastic IKE}), or the
equivalent equations (\ref{IKE-1}) and (\ref{IKE-2}), are formally
similar to the Vlasov equation arising in the kinetic theory of
quasi-linear and strong
turbulence for Vlasov-Poisson plasmas \cite%
{Dupree1966,Weinstock1969,Benford1972} and related renormalized
kinetic theory \cite{Krommes1979}, which are known to lead
generally to a non-Markovian kinetic equation for $\left\langle
f\right\rangle $ alone.
Nevertheless, the stochastic-averaged kinetic equation [\textit{i.e.}, Eq. (%
\ref{IKE-1})] is known to be amenable, under suitable assumptions,
to an approximate Fokker-Planck kinetic equation advancing in time
$\left\langle f\right\rangle $ alone. This is achieved by formally
constructing a perturbative solution of the equation (\ref{IKE-2})
for the stochastic perturbation $\delta f.$ To obtain a convergent
perturbative theory, however, this usually requires the adoption
of a suitable renormalization scheme in order to obtain a
consistent kinetic equation for $\left\langle f\right\rangle .$
This raises the interesting question whether in the framework of
IKT there exist possible alternatives based on the construction of
exact Markovian IKT formulations.

A solution to this problem is provided by the construction of an IKT \emph{%
only for the stochastic-averaged equations} (\ref{8}),
\textit{i.e.}, to be achieved by means of a \emph{Reduced IKT
}\cite{Tessarotto2008z}. Here we
intend to show that this leads to an inverse kinetic equation of the form (%
\ref{Liouville0}),
\begin{equation}
L(\left\langle Z(\mathbf{\alpha })\right\rangle
)f(\mathbf{x},t;\left\langle Z(\mathbf{\alpha })\right\rangle )=0,
\label{Liouville}
\end{equation}%
which is manifestly Markovian
\cite{Tessarotto2008a,Tessarotto2008z} and satisfies also a
constant H-theorem (i.e., it conserves the Boltzmann-Shannon
entropy). Here $f\equiv f(\mathbf{x},t;\left\langle Z\right\rangle
)$ is the Eulerian local pdf for the stochastic INSE problem,
which advances in time the stochastic-averaged fluid fields
$\left\{ \left\langle Z(\mathbf{\alpha })\right\rangle \right\} $
and $L(\left\langle Z(\mathbf{\alpha })\right\rangle )$ is the
corresponding streaming operator, to be defined in terms of the
vector field
\begin{equation}
\mathbf{F}_{1}(\mathbf{x},t;f,\left\langle Z(\mathbf{\alpha
})\right\rangle
)=\left\{ \mathbf{F}(\mathbf{x},t;f,\left\langle Z(\mathbf{\alpha }%
)\right\rangle ),\mathbf{Y}(\mathbf{x},t;f,\left\langle Z(\mathbf{\alpha }%
)\right\rangle )\right\} ,
\end{equation}%
with $\mathbf{F}_{1}$ to be interpreted again as a \emph{mean
field force} acting on a particle with state $\mathbf{x}.$ Besides
the requirement of validity of the fluid equations (\ref{8})
\cite{Ellero2005}, let us impose
that vector $\mathbf{X}$ has the form $\mathbf{X=}\left\{ \mathbf{v,y,F,Y}%
\right\} ,$\textbf{\ }where $\mathbf{F}(\mathbf{x,}t;f,\left\langle Z(%
\mathbf{\alpha })\right\rangle )$ and $\mathbf{Y}(\mathbf{x,}%
t;f,\left\langle Z(\mathbf{\alpha })\right\rangle )$ are both
assumed generally functionally dependent on the local pdf. Again,
by appropriate choice of the mean field forces $\mathbf{F,Y}$, the
moment equations can be proven to satisfy identically the fluid
equations [Eqs.(\ref{8})], as well the appropriate initial and
boundary conditions. Requiring that $f$ is a strictly positive let
us require that there results identically in $\Gamma \times I:$

\begin{equation}
\left\{
\begin{array}{c}
1=\int d\mathbf{v}d\mathbf{y}f(\mathbf{x},t;\left\langle
Z\right\rangle ),
\\
\left\langle \mathbf{V}(\mathbf{r,}t;\mathbf{\alpha })\right\rangle =\int d%
\mathbf{v}d\mathbf{\mathbf{y}v}f(\mathbf{x},t;\left\langle
Z\right\rangle ),
\\
\widehat{p}_{1}(\mathbf{r,}t)=\rho _{o}\int d\mathbf{v}d\mathbf{y}\frac{1}{3}%
\left\langle \mathbf{u}\right\rangle
^{2}f(\mathbf{x},t;\left\langle
Z\right\rangle ) \\
\left\langle \mathbf{B}(\mathbf{r,}t;\mathbf{\alpha })\right\rangle =\int d%
\mathbf{v}d\mathbf{\mathbf{y}y}f(\mathbf{x},t;\left\langle
Z\right\rangle ),
\\
\left\langle S_{T}(t;\mathbf{\alpha })\right\rangle =S(f(\mathbf{x}%
,t;\left\langle Z\right\rangle )).%
\end{array}%
\right.  \label{moments-1}
\end{equation}%
Here $S(f(\mathbf{x},t;\left\langle Z\right\rangle ,\mathbf{\alpha
}))$ is the (Shannon) entropy integral
\begin{equation}
S(f(\mathbf{x},t;\left\langle Z\right\rangle ,\mathbf{\alpha }%
))=-\int_{\Gamma }d\mathbf{x}f(\mathbf{x},t;\left\langle Z(\mathbf{\alpha }%
)\right\rangle )\ln f(\mathbf{x},t;\left\langle Z(\mathbf{\alpha }%
)\right\rangle ),
\end{equation}%
$\mathbf{u=v}-\mathbf{V}(\mathbf{r},t;\mathbf{\alpha })$ is the
relative velocity, while
\begin{equation}
\widehat{p}_{1}(\mathbf{r,}t)=P_{0}(t)+\left\langle p(\mathbf{r,}t;\mathbf{%
\alpha })\right\rangle  \label{kinetic pressure}
\end{equation}%
is\ the kinetic pressure and $P_{0}(t)$ a smooth real function
(pseudo-pressure) to be suitably defined (see below). Again the
form of the local pdf $f\equiv f(\mathbf{x},t;\left\langle
Z\right\rangle )$ can be chosen in such a way to satisfy the
principle of entropy maximization (PEM) \cite{Jaynes1957},
\textit{i.e.}, imposing, at the initial time $t=t_{o},$ the
variational equation $\delta S(f)=0,$ with $\delta ^{2}S(f)<0,$
while
requiring that $f$ belongs to the functional class $\left\{ f(\mathbf{x}%
,t;\left\langle Z(\mathbf{\alpha )}\right\rangle )\right\} ,$
manifestly prescribed by the constraints (\ref{moments-1}). Hence,
PEM implies yields necessarily (for $t=t_{o}$) that the initial
pdf must be of the form
\begin{equation}
f_{M}(\mathbf{x},t;\left\langle Z\right\rangle )=\frac{1}{\pi
^{2}v_{th}^{3}v_{th,T}}\exp \left\{ -X^{2}-Y^{2}\right\} .
\label{Maxwellian}
\end{equation}%
Here, in difference to Eqs.(\ref{pos-1}) and (\ref{pos-2}), $X^{2}$ and $%
Y^{2}$ now denote
\begin{equation}
X^{2}=\frac{\left\langle \mathbf{u}\right\rangle
^{2}}{vth_{p}^{2}},
\end{equation}%
\begin{equation}
Y^{2}=\frac{\left\langle \mathbf{w}\right\rangle
^{2}}{v_{th,T}^{2}},
\end{equation}%
where $\mathbf{u=v-V,}$ $\mathbf{w=y-B,}$
$v_{th}^{2}=2\widehat{p}_{1}/\rho _{o}$ and $v_{th,T}^{2}$ is an
'\textit{a priori}' arbitrary strictly positive constant.
Eq.(\ref{Liouville}) implies the construction of a
suitable classical dynamical system, defined by a phase-space map $\mathbf{x}%
_{o}\rightarrow \mathbf{x}(t)=T_{t,t_{o}}\mathbf{x}_{o},$ where
$T_{t,t_{o}}$ is the evolution operator \cite{Ellero2005}
generated by the stochastic
phase-space Lagrangian equation (\ref{DYN-SYS}) with vector field $\mathbf{X}%
(\mathbf{x},t,\left\langle Z(\mathbf{\alpha })\right\rangle )$ and Jacobian $%
J(\mathbf{x}(t),t,\left\langle Z(\mathbf{\alpha )}\right\rangle
).$

Therefore, the\ Eulerian IKE (\ref{Liouville}) can also be
represented in the Lagrangian form [equivalent to
Eq.(\ref{Liouville})]

\begin{equation}
J(\mathbf{x}(t),t,\left\langle Z(\mathbf{\alpha )}\right\rangle )f(\mathbf{x}%
(t),t;\left\langle Z(\mathbf{\alpha )}\right\rangle )=f(\mathbf{x}%
_{o},t_{o};\left\langle Z_{o}(\mathbf{\alpha })\right\rangle )\equiv f_{o}(%
\mathbf{x}_{o};\left\langle Z_{o}(\mathbf{\alpha })\right\rangle
), \label{Eq.2}
\end{equation}%
(\emph{Lagrangian IKE}), Here, $f_{o}(\mathbf{x}_{o};\left\langle Z_{o}(%
\mathbf{\alpha })\right\rangle )$ is{\ a suitably smooth initial pdf and }%
\begin{equation}
J(\mathbf{x}(t),t;\left\langle Z(\mathbf{\alpha })\right\rangle
)=\left\vert \frac{\partial \mathbf{x}(t)}{\partial
\mathbf{x}_{o}}\right\vert
\end{equation}%
is the Jacobian of the map $\mathbf{x}_{o}\rightarrow
\mathbf{x}(t).$ Then
it is immediate to prove that the stochastic MHD problem admits a \emph{%
Reduced IKT}. \ For the sake of simplicity here we present the
result which holds in the case of a Maxwellian (local) pdf of the
type (\ref{Maxwellian}).

\textbf{Theorem 3 - Markovian Reduced IKT for the
stochastic-averaged MHD equations}

\emph{Let us assume that:} \emph{A}$_{1})$\emph{\ the stochastic
MHD problem (\ref{8}) admits a smooth strong solution in
}$\overline{\Gamma }\times I,$ \emph{with }$I$\emph{\ a finite
time interval}$;$ \emph{A}$_{2}$\emph{) the}
\emph{mean-field force }%
\begin{equation}
\mathbf{F}_{1}(\mathbf{x},t;f,\left\langle Z(\mathbf{\alpha
})\right\rangle
)=\left\{ \mathbf{F}(\mathbf{x},t;f,\left\langle Z(\mathbf{\alpha }%
)\right\rangle ),\mathbf{Y}(\mathbf{x},t;f,\left\langle Z(\mathbf{\alpha }%
)\right\rangle )\right\}  \label{mean-field force average}
\end{equation}%
\emph{\ is defined by }%
\begin{equation}
\mathbf{F}(\mathbf{x},t;f_{M},\left\langle Z(\mathbf{\alpha
})\right\rangle )=\mathbf{F}_{0}+\mathbf{F}_{1},  \label{mean-F-1}
\end{equation}

\begin{equation}
\mathbf{Y(\mathbf{x},}t;f_{M},\left\langle Z(\mathbf{\alpha })\right\rangle )%
\mathbf{=}\left\langle \mathbf{u\cdot \nabla B}\right\rangle
-\left\langle \mathbf{h}\right\rangle -\frac{c}{4\pi \sigma
}\nabla ^{2}\left\langle \mathbf{B}\right\rangle
\mathbf{+}\left\langle \mathbf{w}\nabla \cdot
\mathbf{B}\right\rangle \mathbf{,}  \label{mean-F-2}
\end{equation}

\emph{where }$\mathbf{F}_{0}$ \emph{and}\textbf{\
}$\mathbf{F}_{1}$\emph{\
read respectively }%
\begin{equation}
\mathbf{F}_{0}\mathbf{(x,}t;f_{M},\left\langle Z(\mathbf{\alpha }%
)\right\rangle )=\frac{1}{\rho _{o}}\left\langle \mathbf{f}\right\rangle +%
\frac{1}{2}\left\langle \mathbf{u}\cdot \nabla
\mathbf{V}\right\rangle
\mathbf{+}\frac{1}{2}\left\langle \mathbb{\nabla }\mathbf{V\cdot u}%
\right\rangle \mathbf{+}\nu \nabla ^{2}\left\langle
\mathbf{V}\right\rangle \mathbf{,}
\end{equation}%
\begin{equation}
\mathbf{F}_{1}\mathbf{(x,}t;f_{M},\left\langle Z(\mathbf{\alpha }%
)\right\rangle )=\frac{\left\langle \mathbf{u}\right\rangle }{2}\frac{1}{%
\widehat{p}_{1}}+\frac{v_{th}^{2}}{2}\nabla \ln p_{1}\left\{ \frac{%
\left\langle \mathbf{u}\right\rangle
^{2}}{v_{th}^{2}}-\frac{3}{2}\right\} ,
\end{equation}%
\emph{while }%
\begin{equation}
A\equiv \frac{\partial }{\partial t}\left( P_{0}(t)+\left\langle
p\right\rangle \right) +\left\langle \mathbf{V}\right\rangle
\mathbf{\cdot \nabla }\left( P_{0}(t)+\left\langle p\right\rangle
\right)
\end{equation}%
\emph{is prescribed by the requirement of validity of the
stochastic-averaged energy equation \cite{Tessarotto2008a};} \emph{A3) }$f(%
\mathbf{x,}t;\left\langle Z(\mathbf{\alpha })\right\rangle )$ \emph{%
satisfies suitable initial and boundary condition consistent with
the
initial-boundary value problem} \emph{\ (\ref{8}) (see Ref. \cite{Ellero2005}%
); A4) the initial local pdf, }$f(\mathbf{x,}t_{o};\left\langle Z_{o}(%
\mathbf{\alpha })\right\rangle ),$ \emph{is suitably smooth an
strictly positive}. \emph{\ }\emph{It follows that:}\emph{\ }

\emph{1) the velocity-moment equations of IKE (\ref{Liouville})
evaluated for the weight functions
}$G(\mathbf{x},t)=v,\frac{1}{3}u^{2}$\emph{\ coincide with the
stochastic INSE equations); }

\emph{2)} \emph{if }$f(t_{o},\left\langle Z(\mathbf{\alpha })\right\rangle )$%
\emph{\ is strictly positive so it is }$f(t_{o},\left\langle Z(\mathbf{%
\alpha })\right\rangle )$ \emph{in the whole domain} $\overline{\Gamma }%
\times I;$

\emph{3) the Maxwellian local pdf (\ref{Maxwellian}) is a
particular solution of IKE if an only if the fluid fields
}$\left\{ \left\langle Z\right\rangle \right\} $\emph{\ satisfy
the stochastic-averaged MHD problem (\ref{8}); }

\emph{4) the pseudo-pressure }$P_{o}(t)$ \emph{can be uniquely
determined in the time interval }$I$ \emph{in such a way that,
denoting }$f(t,\left\langle Z(\mathbf{\alpha })\right\rangle
)\equiv f(\mathbf{x},t;\left\langle Z\right\rangle ),$\emph{\ in
the same time interval }$I$ \emph{there results
identically }%
\begin{equation}
\frac{\partial }{\partial t}S(f(t,\left\langle Z(\mathbf{\alpha }%
)\right\rangle ))=0  \label{H-theorem}
\end{equation}%
\emph{(constant H-theorem).}

\noindent \emph{PROOF}

The proof is immediate for proposition 1), while 2) follows
invoking the
Lagrangian IKE (\ref{Eq.2}) and by noting that by construction $J(\mathbf{x}%
(t),t;\left\langle Z(\mathbf{\alpha })\right\rangle )>0$ in the
whole domain $\overline{\Gamma }\times I$. To prove 3) let us
assume that a strong solution of the stochastic INSE problem
exists. In such a case it is immediate to prove that
$f_{M}(\mathbf{x}(t),t;\left\langle Z\right\rangle )$ is a
particular solution of the inverse kinetic equation
(\ref{Liouville})$.$
In fact, substituting (\ref{Maxwellian}) in the inverse kinetic equation (%
\ref{Liouville}) it follows:
\begin{eqnarray*}
&&\left. Lf_{M}(\mathbf{x},t;\left\langle Z(\mathbf{\alpha
})\right\rangle )=\left\{ \frac{\partial }{\partial t}\left\langle
\mathbf{V}\right\rangle \mathbf{+v\cdot \nabla }\left\langle
\mathbf{V}\right\rangle \right\}
\mathbf{\cdot }\frac{\left\langle \mathbf{u}\right\rangle \rho _{o}}{%
\widehat{p}_{1}}f_{M}\mathbf{+}\right. \\
&&+\left\{ \frac{\partial }{\partial t}\left\langle
\mathbf{B}\right\rangle \mathbf{+v\cdot \nabla }\left\langle
\mathbf{B}\right\rangle \right\}
\mathbf{\cdot }\frac{2\left\langle \mathbf{w}\right\rangle }{v_{th,T}^{2}}%
f_{M}
\end{eqnarray*}%
\begin{equation}
\left. +\left\{ \frac{\partial }{\partial t}\ln \widehat{p}_{1}\mathbf{%
+v\cdot \nabla }\ln \widehat{p}_{1}\right\} \left\{
\frac{\left\langle u\right\rangle
^{2}}{v_{th}^{2}}-\frac{3}{2}\right\} f_{M}-\right.
\end{equation}%
\[
+\left\{ -\mathbf{F}\cdot \frac{\left\langle
\mathbf{u}\right\rangle \rho
_{o}}{\widehat{p}_{1}}-\mathbf{Y}\cdot \frac{2\left\langle \mathbf{w}%
\right\rangle }{v_{th,T}^{2}}+\frac{\partial }{\partial
\mathbf{v}}\cdot \mathbf{F}+\frac{\partial }{\partial
\mathbf{y}}\cdot \mathbf{Y}\right\} f_{M},
\]%
which manifestly implies Eqs.(\ref{8}). \ Instead, if we assume that in $%
\Gamma \times I,$ $f\equiv f_{M}(\mathbf{x},t;\left\langle Z(\mathbf{\alpha }%
)\right\rangle )$ is a particular solution of the inverse kinetic
equation
(IKE) (\ref{Liouville}), which fulfills identically the constraint equation (%
\ref{H-theorem}), thanks to proposition 1) it follows that the fluid fields $%
\left\langle \mathbf{V}\right\rangle ,\left\langle p\right\rangle
$ are solutions of the INSE equations. Finally to prove
proposition 4) and 5), let us invoke IKE to evaluate the entropy
production\ rate, which reads
\begin{equation}
\frac{\partial }{\partial t}S(f(t,\left\langle Z(\mathbf{\alpha }%
)\right\rangle ))=-\int_{\Gamma
}d\mathbf{x}f(\mathbf{x,}t;\left\langle
Z\right\rangle )\left[ \frac{\partial }{\partial \mathbf{v}}\cdot \mathbf{F}(%
\mathbf{x},t;f)+\frac{\partial }{\partial \mathbf{y}}\cdot \left\{ \mathbf{Y}%
(\mathbf{x},t;f)\right\} \right] .
\end{equation}%
Hence, in a bounded time interval $I$ it always possible to
satisfy the
constraint placed by the constant H-theorem, requiring%
\begin{eqnarray}
&&\left. \frac{\partial }{\partial t}S(f(t,\left\langle Z(\mathbf{\alpha }%
)\right\rangle ))=\right.  \label{constant H-theorem-2} \\
&=&-\frac{3}{2}\int_{\Omega
}d^{3}\mathbf{r}\frac{1}{P_{0}(t)+\left\langle p\right\rangle
}\left[ \frac{\partial }{\partial t}\left(
P_{0}(t)+\left\langle p\right\rangle \right) +\left\langle \mathbf{V}%
\right\rangle \mathbf{\cdot \nabla }\left( P_{0}(t)+\left\langle
p\right\rangle \right) \right] =0.
\end{eqnarray}%
This delivers an ordinary differential equation for the pseudo-pressure $%
P_{0}(t)$. Assuming that the fluid fields $\left\langle \mathbf{V}%
\right\rangle ,\left\langle p\right\rangle $ are suitably smooth,
this equation can always be fulfilled at least in the case in
which the domain of existence $I$ is a finite time interval$.$
Q.E.D.

Let us briefly analyze the consequences of the theorem, with
particular reference to the condition of strict positivity of the
pdf [proposition 2)] and the constant H-theorem [Eq.(\ref{constant
H-theorem-2}), see proposition 5)].\ It is well-known that
H-theorem \ assures, for arbitrary (and suitably smooth) initial
conditions of the pdf, also the strict positivity of the kinetic
distribution function. Nevertheless, in view of the validity of
proposition 2), the requirement of its validity is, in principle,
unnecessary. In fact, the H-theorem is only necessary because of
the requirements posed by the condition of isentropic flow [see
Eqs.(\ref{8})] and the correspondence principle [defined by
Eqs.(\ref{moments-1})]. However, there is (another) important
physical consequence of the theorem.
In fact, if the requirement is posed that the initial pdf $%
f(t_{o},\left\langle Z(\mathbf{\alpha })\right\rangle )$ satisfies
PEM, i.e., it maximizes $S(f(t_{o},\left\langle Z(\mathbf{\alpha
})\right\rangle )),$ the constant H-theorem assures that in the
the domain of existence $I,$ $f(t,\left\langle Z(\mathbf{\alpha
})\right\rangle )$ must necessarily maximize $S(f(t,\left\langle
Z(\mathbf{\alpha })\right\rangle )).$
Therefore, if for all $t\in I,$ $f(t,\left\langle Z(\mathbf{\alpha }%
)\right\rangle )$ is assumed to belong to the same functional
class specified above $\left\{ f(t,\left\langle Z(\mathbf{\alpha
})\right\rangle
)\right\} ,$ then necessarily $f(t,\left\langle Z(\mathbf{\alpha }%
)\right\rangle )$ is a kinetic equilibrium, to be identified with
the local Maxwellian distribution (\ref{Maxwellian}). Similar
conclusions apply manifestly to THM.2 too, provided the functional
class is suitably defined
(see THM2.) and the kinetic equilibrium is identified with ((\ref%
{Maxwellian0})).

Another interesting issue is provided by the comparison between
the two stochastic IKT-approaches discussed above, i.e., the
Complete and the Reduced IKT's here developed. Key differences
between them are manifestly provided both by the form of the
inverse kinetic equations and the choice of the initial conditions
for the corresponding pdf. This is due to the different definition
of the mean field force in the two case3s, given respectively by
Eqs.(\ref{mean-field-force- stochastic}),(\ref{mean-field force
average}) and moreover by the different requirements posed in the
two
case by PEM (the principle of entropy maximization; Janyes, 1957 \cite%
{Jaynes1957}). In fact, the initial pdf $f(t_{o},\left\langle Z(\mathbf{%
\alpha })\right\rangle ),$ which in both cases is an extremal of
the Boltzmann-Shannon entropy, namely is such that $\delta
S(f(t_{o},\left\langle Z(\mathbf{\alpha })\right\rangle ))=0,$
belongs to
different functional classes $\left\{ f(t_{o},\left\langle Z(\mathbf{\alpha }%
)\right\rangle )\right\} $. The precise definition of $\left\{
f(t_{o},\left\langle Z(\mathbf{\alpha })\right\rangle )\right\} $
in the two cases is actually important since it determines
uniquely the initial
condition [see related discussion in Refs \cite{Tessarotto2008z} and\ \cite%
{Tessarotto2008z2}]. Provided either the total initial fluid fields $Z(%
\mathbf{r,}t_{o};\mathbf{\alpha })\mathbf{=}Z_{o}(\mathbf{r};\mathbf{\alpha }%
)$ or only the stochastic averaged ones $\left\langle Z(\mathbf{r,}t_{o};%
\mathbf{\alpha })\right\rangle \mathbf{=}\left\langle Z_{o}(\mathbf{r};%
\mathbf{\alpha })\right\rangle $ are prescribed

In both cases [see THM's 1, 2 and 3] the initial kinetic
probability density is provided\ by a Maxwellian-type kinetic
equilibrium distribution. This is found as the "most-likely"
initial kinetic probability density at $t=t_{o},$
in agreement with PEM. In particular, in the first case the distribution $%
f_{M}(\mathbf{x},t;Z)$ [see Eq.(\ref{Maxwellian0})] follows by
requiring that\ $\left\{ f(t_{o},\mathbf{\alpha })\right\} $ is
prescribed imposing
the constraints defined by Eqs.(\ref{moments-0}) - to be evaluated at $%
t=t_{o}$ - while in the second $f_{M}(\mathbf{x},t;\left\langle
Z\right\rangle )$ [see Eq.(\ref{Maxwellian})] follows from the constraints (%
\ref{moments-1}). This means that in the first case the initial
fluid fields the compete set of fluid fields $\left\{ Z\right\}
_{t=t_{o}}$ must be considered as prescribed, while in the second
only the stochastic-averaged fluid fields $\left\{ \left\langle
Z\right\rangle \right\} _{t=t_{o}}$ is initially known (so that in
this case the initial value of the stochastic fluctuations
$\left\{ \delta Z\right\} _{t=}$ remains in this case arbitrary).

\section{Generalizations to nonhomogeneous and nonstationary turbulence}

Turbulence models depend both on the definitions of the hidden variables $%
\mathbf{\alpha }$ and of the related rs-pdf $g$ (see also previous section)$%
. $ This involves specifically the case in which $g$ is taken of the form $g(%
\mathbf{r},t;\mathbf{\alpha }),$ which corresponds to the
assumption of nonhomogeneous and nonstationary turbulence. It is
obvious that, unless specific additional assumptions are
introduced, such models are non-unique, due to the arbitrariness
of their possible choices [both for $\mathbf{\alpha }$ and $g$].
Their unique determination, however, may in principle be achieved
either based on suitable '\textit{ad hoc}' mathematical models or
the phenomenology of (magneto-)fluids, based either on first principles (%
\textit{i.e.}, implied by the underlying microscopic molecular
dynamics and statistical mechanics), or the observation of real
fluids or obtained from numerical experiments. Nevertheless,
despite this arbitrariness, in the
particular case in which the functional form of the fluid equations [\textit{%
i.e.}, Eqs.(\ref{(1)})] remains unchanged, the generalization of
the present theory is still possible. This requires solely
suitable smoothness assumptions to be satisfied by the rs-pdf. On
the contrary, a similar
conclusion is not obvious if the form of the fluid equations [Eqs.(\ref{(1)}%
)] is modified. In fact, in such a case, \textit{i.e.}, for
arbitrary
turbulence-modified fluid equations, the existence of an IKT-approach '%
\textit{a priori}' cannot generally be assured.

\section{Concluding remarks}

A statistical model of MHD turbulence has been pointed out
utilizing the inverse kinetic theory (IKT)\ recently developed for
classical and quantum
fluids \cite%
{Ellero2000,Tessarotto2004,Ellero2005,Tessarotto2006,Tessarotto2007a}.
Basic feature of the theory is that the IKT has been constructed
in such a way to satisfy exactly the stochastic MHD equations
describing an incompressible, viscous, quasi-neutral, isentropic,
isothermal and resistive magnetofluid. As consequence an inverse
kinetic equation has been determined which uniquely advances in
time the local pdf. \

The main result of the paper is represented by THM.2 and 3,
yielding IKT-approaches to the stochastic MHD equations and
particularly by the inverse kinetic equations [\textit{i.e.},
Eqs.(\ref{IKE-1}),(\ref{IKE-2}) and (\ref{Liouville})], which
advance in time respectively the stochastic or the
stochastic-averaged local pdf. The first two equations
[\textit{i.e.}, Eqs.(\ref{IKE-1}) and (\ref{IKE-2})], determine
uniquely the evolution of
the stochastic fluid fields, \textit{i.e.}, both the stochastic averages $%
\left\{ \left\langle Z(\mathbf{\alpha })\right\rangle \right\} $
and the
corresponding stochastic fluctuations $\left\{ \delta Z(\mathbf{\alpha }%
)\right\} .$ Instead, Eq.(\ref{Liouville}) determines only
$\left\{ \left\langle Z(\mathbf{\alpha })\right\rangle \right\} $
(\textit{i.e.}, provided $\left\{ \delta Z(\mathbf{\alpha
})\right\} $ is known). In both cases the inverse kinetic
equations have the property that their velocity-moments (to be
defined in terms of appropriate weighted velocity-space integrals)
satisfy identically a closure condition. In fact, there exists, by
construction, a subset of the moment equations which is closed.
Such a set is defined so that it coincides with the appropriate
stochastic fluid equations [\textit{i.e.}, either Eqs.(\ref{8}) or (\ref{8b}%
)].

The theory displays several interesting new features, with respect
to customary phase-space statistical approaches to turbulence (see
for example \cite{Monin1975,Pope2000}). In particular:

\begin{enumerate}
\item it is based on the introduction of a local position-velocity joint
probability density function (local pdf), rather than the usual
velocity-difference pdf (which typically affords only an
approximate description);

\item Eulerian and Lagrangian phase-space formulations are manifestly
equivalent;

\item a remarkable, and in a sense, surprising feature of the present theory
is that the equation advancing in time the stochastic-averaged pdf
is can also be set in the form of a Markovian kinetic equation
\cite{Dupree1972}. This feature is actually "built in" the IKT
approach, thanks to the definition here adopted for the time
evolution operator $T_{t,t_{o}}$ \ which advances in time the
local pdf (and hence the same fluid fields).
\end{enumerate}

In our view the IKT-theory here presented is a useful setting for
the investigation of theoretical and mathematical aspects of
turbulence phenomena which may potentially occur in a variety of
fluids (both incompressible and compressible) and in particular
magnetofluids. Several interesting applications and
generalizations of the theory are possible, which include - for
example - the investigation of turbulence in fluid mixtures and in
thermo-magnetofluids. The proper treatment of such phenomena
requires a consistent description of phase-space dynamics, to be
based on the IKT approach developed in this paper. We remark, in
this connection,
that in the present approach both the specification of the hidden variables $%
\mathbf{\alpha }$ and of the reduced stochastic probability
density ($g$) remains in principle completely arbitrary. This
feature is potentially important since it permits, in principle,
the systematic treatment of all the possible different sources of
stochasticity here pointed out, a problem which remains still
fundamentally unsolved to date. Finally, the extension of the
theory to nonhomogeneous and nonstationary turbulence has been
pointed out (Sec.7). \ Related developments will be discussed in
greater detail elsewhere.

% ------------------------------------ Acck --------------------------------

\section*{Acknowledgements}

$^{\S }$ Work developed in the framework of the MIUR (Italian
Ministry of University and Research) PRIN Research Program
\textquotedblleft Modelli della teoria cinetica matematica nello
studio dei sistemi complessi nelle scienze
applicate\textquotedblright , the European COST action P17 and the
European Research Network on Magnetoscience (GDRE GAMAS). The
partial support of the GNFM (National Group of Mathematical
Physics) of INDAM (National Institute of Advanced Mathematics,
Italy) is acknowledged.

\bigskip

\section{Appendix: stochastic and deterministic functions}

A possible convenient definition of stochastic function $A$ (to be
identified with the generic fluid field $Z_{i}$) is the following
one:

\textbf{Definition - Stochastic function \label{Definition}}

{\emph{A }function $A$ is denoted as stochastic on the set } $\overline{%
\Omega }\times \overline{I}\times V_{\alpha }${\ if: }

1) $A \equiv A(\mathbf{r},t,\mathbb{\alpha })$ is assumed to be a
real function defined on the set $\overline{\Omega }\times
\overline{I}\times V_{\alpha }$ which depends on the real
hidden-variables (to be denoted as \emph{stochastic parameters})
$\mathbf{\alpha }\in V_{\alpha }\subseteq \mathbb{R} ^{k};$

2) the $\alpha ^{\prime }$s are characterized by a probability
density $g,$ denoted as \emph{reduced stochastic probability
density function }(rs-pdf)
defined on $V_{\alpha }$ , with $V_{\alpha }$ a non-empty subset of $%
\subseteq \mathbb{R}^{k}$ and $k\geq 1$. Hence, $g$ is necessarily
non-negative ($g\geq 0$) [or even strictly positive ($g>0)$] and such that%
\begin{equation}
\int\limits_{V_{\alpha }}d\mathbb{\alpha }g=1,
\label{nromalization}
\end{equation}%
$d\mathbf{\alpha }$ denoting the canonical measure on $V_{\alpha
};$

3) introducing the \emph{stochastic averaging operator}
\begin{equation}
\left\langle \cdot \right\rangle \equiv \int\limits_{V_{\alpha }}d\mathbb{%
\alpha }g\cdot ,  \label{stochastic averaging operator}
\end{equation}%
where the integration is performed at constant $\left(
\mathbf{r},t\right)
\in $ $\overline{\Omega }\times \overline{I},$ the stochastic function $A(%
\mathbf{r},t,\mathbb{\alpha })$ is assumed to admit everywhere in $\overline{%
\Omega }\times \overline{I}$ the \emph{stochastic average}
\begin{equation}
\left\langle A\right\rangle \equiv \int\limits_{V_{\alpha }}d\mathbb{\alpha }%
gA(\mathbf{r},t,\mathbb{\alpha });  \label{stochastic agerage}
\end{equation}

In contrast, a deterministic function can be defined as:

\textbf{Definition - Deterministic function}

A {function $A(\mathbf{r},t,\mathbb{\alpha })$ is denoted as \emph{%
deterministic} if there results identically on the set $\overline{\Omega }%
\times V_{\alpha }\times \overline{I}$
\begin{equation}
A=\left\langle A\right\rangle .
\end{equation}%
} This requires that that either $A$ is independent of the parameters $%
\mathbf{\alpha }$ or the probability density $g$ is a
$k$-dimensional Dirac delta, \textit{i.e.}, it is of the form
$g=\delta ^{(k)}(\mathbf{\alpha -\alpha }_{o})\equiv
\prod\limits_{i=1,k}\delta (\alpha _{i}-\alpha _{oi}),$ with
$\mathbf{\alpha }_{o}$ a prescribed element of $V_{\alpha }$.

% ------------------------------------ Bib --------------------------------

% \bibliographystyle{mhd}
% \bibliography{thesis}

\begin{thebibliography}{99}
\bibitem{Theory} Turbulence theory involves therefore the formulation and
investigation of stochastic models of fluids.
\bibitem{Kinetic equilibrium} By definition the probability density [$f(%
\mathbf{x},t;Z(\mathbf{\alpha }),\mathbf{\alpha })$ or $f(\mathbf{x}%
,t,\left\langle Z(\mathbf{\alpha })\right\rangle )$] is a kinetic
equilibrium if it depends on time only at most via the fluid
fields
[respectively $Z(\mathbf{\alpha })$ or $\left\langle Z(\mathbf{\alpha }%
)\right\rangle $].
\end{thebibliography}

\end{document}